\documentclass[twocolumn,aps,pre,showpacs,superscriptaddress]{revtex4}
\usepackage{epsfig,graphicx}
\usepackage{bm}
\usepackage{amsmath}
\usepackage{times}

\begin{document}

\title{Controlling collective dynamics in complex, minority-game resource-allocation systems}

\author{Ji-Qiang Zhang} \affiliation{Institute of
Computational Physics and Complex Systems, Lanzhou University,
Lanzhou Gansu 730000, China}

\author{Zi-Gang Huang}\email{huangzg@lzu.edu.cn} \affiliation{Institute of
Computational Physics and Complex Systems, Lanzhou University,
Lanzhou Gansu 730000, China}\affiliation{School of Electrical,
Computer, and Energy Engineering, Arizona State University, Tempe,
AZ 85287, USA}

\author{Jia-Qi Dong} \affiliation{Institute of
Computational Physics and Complex Systems, Lanzhou University,
Lanzhou Gansu 730000, China}

\author{Liang Huang}
\affiliation{Institute of Computational Physics and Complex
Systems, Lanzhou University, Lanzhou Gansu 730000,
China}\affiliation{School of Electrical, Computer, and Energy
Engineering, Arizona State University, Tempe, AZ 85287, USA}

\author{Ying-Cheng Lai} \affiliation{School of Electrical,
Computer, and Energy Engineering, Arizona State University, Tempe,
AZ 85287, USA} \affiliation{Department of Physics, Arizona State
University, Tempe, AZ 85287, USA}

\date\today

\begin{abstract}
Resource allocation takes place in various kinds of real-world
complex systems, such as the traffic systems, social services
institutions or organizations, or even the ecosystems. The
fundamental principle underlying complex resource-allocation
dynamics is Boolean interactions associated with {\em minority
games}, as resources are generally limited and agents tend to
choose the least used resource based on available information. A
common but harmful dynamical behavior in resource-allocation
systems is herding, where there are time intervals during which a
large majority of the agents compete for a few resources, leaving
many other resources unused. Accompanying the herd behavior is
thus strong fluctuations with time in the number of resources
being used. In this paper, we articulate and establish that an
intuitive control strategy, namely pinning control, is effective
at harnessing the herding dynamics. In particular, by fixing the
choices of resources for a few agents while leaving majority of
the agents free, herding can be eliminated completely. Our
investigation is systematic in that we consider random and
targeted pinning and a variety of network topologies, and we carry
out a comprehensive analysis in the framework of mean-field theory
to understand the working of control. The basic philosophy is then
that, when a few agents waive their freedom to choose resources by
receiving sufficient incentives, majority of the agents benefit in
that they will make fair, efficient, and effective use of the
available resources. Our work represents a
basic and general framework to address the fundamental issue of
fluctuations in complex dynamical systems with significant
applications to social, economical and political systems.
\end{abstract}

\maketitle

\section{Introduction} \label{sec:intro}

Resource allocation is an essential process in many kinds of
real-world systems, such as traffic systems (e.g., Internet, urban
traffic grids, rail and flight networks), social service
institutions or organizations (e.g., schools, marts, banks, and
financial markets), and ecosystems of various sizes. The
underlying system typically contains a large number of interacting
components or agents on a hierarchy of scales, and there are
multiple resources available for each agent. As a result, complex
behaviors are expected to emerge ubiquitously in the dynamical
process of resource allocation. In a typical situation, agents or
individuals possess similar capabilities, who share the common
goal of pursuing as high payoffs as possible. To exploit the
resource allocation dynamics in multi-agent systems to reduce the
likelihood of or even to eliminate harmful or catastrophic
behaviors is of significant interest.

A general framework to address and understand the extremely rich
and complex dynamics of many real-world systems is complex
adaptive systems \cite{Kauffman:book,Levin:1998,AAP:book}.
Especially suitable for resource-allocation dynamics is the
paradigm of minority-game (MG) dynamics \cite{CZ:1997}, introduced
by Challet and Zhang to address the classic El Farol
bar-attendance problem conceived by Arthur \cite{Arthur:1994}. In
an MG system, each agent makes choice ($+1$ or $-1$, e.g., to
attend a bar or to stay at home) based on available global
information in the memory such as the winning choice in a previous
round of interaction. In particular, the agents who got the
minority choice are rewarded, and those belonging to the majority
group lose due to limited resources. The MG dynamics has
been studied extensively in the last decade or so
\cite{CM:1999,CMZ:2000,MMM:2004,
BMM:2007,RMR:1999,EZ:2000,KSB:2000,Slanina:2000,ATBK:2004,JHH:1999,HJJH:2001,
LCHJ:2004,TCHJ:2005,CMM:2008,BMFM:2008,XWHZ:2005a,ZZZH:2005,MR:2002,Weblink}.

There are two basic and related approaches to the MG problem. One
is based on the mean-field approximation, which was mainly
developed by the statistical-physics community to relate the MG
problem to those associated with non-equilibrium phase transitions
\cite{Moro:2004,CMZ:book,YZ:2008}. Another approach is based on
Boolean-game (BG) dynamics, where for any agent, detailed
information about agents that it interacts with is assumed to be
available, and the agent responds accordingly
\cite{PBC:2000,Vazquez:2000,GL:2002,ZWZYL:2005,
HWGW:2006,HZDHL:2012}. One interesting result was that {\em
coordination} can emerge from local interactions in BG and, as a
result, the system as a whole can achieve ``better than random''
performance in terms of utilization of resources.

A common behavior in many social, economical and ecosystems is
{\em herding}, where many agents take on the same action
\cite{DIMCCWK:2008}. In the past, the herd behavior has been
extensively studied and recognized to be one important factor
contributing to the origin of complexity, which can lead to
enhanced fluctuations and significant reduction in the payoff of
the entire system \cite{EZ:2000,LK:2004,WYZJZW:2005,ZYZXLW:2005}.
For the resource-allocation problem, the desired performance is
that all the resources are used efficiently. When herding occurs,
many agents may go after a very limited number of resources,
causing crowding in the use of these resources, while many other
resources are significantly under-used. The herd behavior is thus
regarded as harmful for resource-distribution systems. An
outstanding issue is whether effective control strategy can be
developed to prevent herding in multi-agent systems with
competition for multiple resources.

In this paper, we investigate a realistically feasible control
strategy to harnessing herding in complex resource-allocation
systems, {\em pinning} control in the framework of Boolean
dynamics. In particular, we show that even a small amount of
pinning can effectively prevent or greatly mitigate the herd
behavior in resource-allocation systems. Take the urban traffic
system as an example.
The basic idea of pinning control is to select certain individuals and pin (or fix) their options to access resources by certain incentives, e.g., compensations or rewards. This is similar
in spirit to the strategy of immunization to prevent wide spread
of disease or virus in complex social or technological networked
systems \cite{CHbA:2003,AM:book,HY:book,PSV:2001,LM:2001}, where
certain individuals are preferred to be immunized to the virus of
concern. However, as we show analytically and demonstrate
numerically in this work, the dynamical mechanism of pinning
control in resource-allocation systems is quite different from
that underlying the immunization problem in complex networks. In
general, we anticipate pinning control to be an effective strategy
to eliminate or suppress harmful herd behaviors in complex systems
describable by Boolean-game dynamics.

In Sec.~\ref{sec:model}, we describe our Boolean-game model under
pinning control. In Sec.~\ref{sec:theory_free_system}, we present
a conventional mean-field theory to analyze the dynamics of free
systems in the absence of control. In
Sec.~\ref{sec:theory_pinning_control}, we point out the
difficulties associated with the conventional mean-field theory
and develop a modified mean-field theory to understand the system
behavior under pinning control. Different pinning schemes and
network topologies are considered. In Sec.~\ref{sec:conclusion},
we offer concluding remarks and discuss relevance of our results
to real-world complex systems.

\section{Model} \label{sec:model}

\subsection{Boolean-game dynamics} \label{subsec:BGD}

Similar to the MG dynamics, there are two alternative resources:
$r=+1$ and $-1$ in a BG dynamic system, and only the agents
belonging to the \emph{global minority} group are rewarded by
$+1$. As a result, the system profit is equal to the number of
agents in the global-minority group. In particular, we consider a
BG dynamic system composed of $N$ agents competing for the two
resources, both of which have accommodating capacity $N/2$. If the
number of agents choosing one given $r$ ($+1$ or $-1$) is smaller
than $N/2$, then it is the global-minority group, and the system
profit is equal to the number of agents in this group.

While, a unique feature of the BG dynamic system, in contrast to
the original MG dynamic system, is that agents make use of only
local information from immediate neighbor in making choice. The
neighborhood of agents is determined by the connecting structure
of the underlying network. Each agent receives inputs from its
neighboring agents and updates its state according to the Boolean
function, a function that generates either $+1$ and $-1$ from the
inputs \cite{GL:2002}. Here, to be concrete, we assume that, an
agent $i$ who has $k_i$ neighbors, will choose $+1$ at time step
$t+1$ with the probability,
\begin{equation} \label{eq:P1}
P_{i\rightarrow\oplus}=n^{t}_{-}/(n^{t}_{+}+n^{t}_{-})=n^{t}_{-}/{k_{i}},
\end{equation}
and $-1$ with the probability
$P_{i\rightarrow\ominus}=1-P_{i\rightarrow\oplus}$. Here,
$n^{t}_{+}$ and $n^{t}_{-}$ respectively are the numbers of $+1$
and $-1$ neighbors of agent $i$ at time step $t$. Notably, agent
in BG dynamics attempts to take on a global-minority choice
without any global information (e.g., previous global-minority
choice), but basing her choice on the observation of neighbors'
previous behavior.

The dynamical variable of the BG system is
$A_t$, the number of $+1$ agents in the system at time step $t$.
Obviously, the optimal solution for the resource allocation is
$A_t=N/2$. A measure of BG system's performance is the
variance of $A_t$:
\begin{equation} \label{eq:sigma}
\sigma^{2}=\frac{1}{T}\sum^{T}_{t=1}(A_{t}-\frac{N}{2})^{2},
\end{equation}
which characterizes the statistical deviation from the optimal
resource utilization over time interval $T$ \cite{ZWZYL:2005}. A
smaller value of $\sigma^{2}$ corresponds to more optimal resource
allocation and thus leads to higher efficiency. A general
phenomenon in BG system is that, as agents strive to join the
minority group, harmful herd behavior can emerge, associated with
which large oscillation in $A_{t}$ takes place. Our goal is to
develop an efficient control strategy to suppress or eliminate the harmful
herd behavior.

\subsection{Pinning control scheme} \label{subsec:pinning_scheme}

Our basic idea to control herd behavior is to ``pin'' certain
agents to freeze their states so as to realize optimal resource
allocation, following the general principle of pinning control of
complex dynamical networks
\cite{WC:2002,LWC:2004,CLL:2007,XLCCY:2007,TWF:2009,PF:2009,YCL:2009}.
In our approach, the fraction of agents to be pinned (fixed) is
$\rho_{pin}$, and the fraction of unpinned or free nodes is
$\rho_{free}=1-\rho_{pin}$. The numbers of free agents and pinned
agents are $N_{f}=N\cdot \rho_{free}$ and $N_{p}=N\cdot
\rho_{pin}$, respectively. The free agents make choices according
to local information, while the inputs from the pinned agents are
fixed.

Our pinning scheme has two features: order of pinning and pinning
pattern. First, the order of pinning denotes the way how certain
agents are chosen for pinning. We consider two methods: random
pinning (RP), where a number of agents are randomly chosen to be
pinned, and degree-preferential pinning (DPP) in which agents are
selected for pinning according to their connectivity or degree in
the underlying network. In particular, agents with higher degree
are more likely to be pinned. These two methods thus correspond to
random error and intentional attack in the literature on
robustness of network systems
\cite{CEbAH:2000,AJB:2000,CNSW:2000,CEbAH:2001}. The second
feature, pinning pattern, defines the particular states that the
selected agents are pinned to. Here we define ``All $+1$'' (or
``All $-1$'') as the pattern where all the pinned agents are
forced to choose $+1$ (or $-1$), and ``Half $\pm 1$'' as the
situation where the agents are to be pinned at $+1$ and $-1$
alternately. The effect of pinning also depends on the network
topology. We consider four representative network topologies:
all-to-all coupling, random \cite{ER:1960}, scale-free
\cite{BA:1999}, and assortatively mixed scale-free networks
\cite{Newman:2002}.

To facilitate a comparative analysis between the free and the
pinned systems, we define a modified cumulative variance as,
\begin{equation} \label{eq:sigmaM}
\sigma^2=\frac{1}{T}\frac{\sum_{t=1}^{T}(A_{t}-\frac{N}{2})^2}{1-\rho_{pin}},
\end{equation}
so that the fluctuations of the systems are comparable with
respect to $\rho_{pin}\in[0,1)$.

\subsection{Simulation results} \label{sec:simulation}

\begin{figure}
\begin{center}
\epsfig{figure=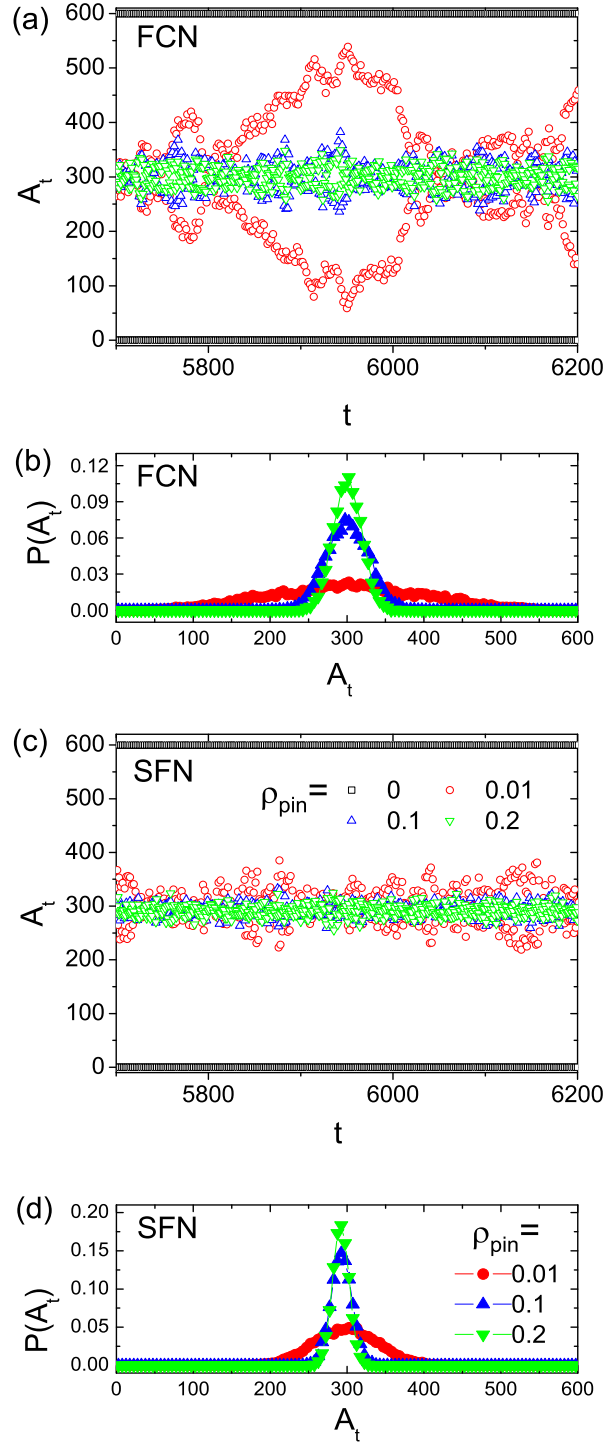,width=\linewidth} \caption{(Color
online). For FCN (a) and SFN network (c), time series $A_t$ for
$\rho_{pin}=0, 0.01$, $0.1$, and $0.2$. The network size is $N=600$
and average degree is $\langle k\rangle=6$. The pinning rule is
DPP with ``Half $\pm 1$.'' The probability density distributions
$P(A_{t})$ for
$\rho_{pin}=0.01$, $0.1$, and $0.2$ (b,d) are calculated from time series $A_t$ of length
$T=10^4$.} \label{fig:At}
\end{center}
\end{figure}

Simulations are carried out for resource-allocation dynamics on
the following types of networks: fully connected networks (FCN),
ER random networks \cite{ER:1960}, scale-free networks (SFN)
\cite{BA:1999}, and under the two pinning schemes (RP or DPP). The
states of the pinned agents are set according to ``Half $\pm 1$'',
``All $+1$'', and ``All $-1$.'' For all the free agents, $+1$ and
$-1$ are uniformly distributed initially. The evolutionary time is
set to be $T=10^{4}$. As an example, Fig.~\ref{fig:At} shows, for
FCN and SFN, time series of $A_{t}$ for different pinning fraction
$\rho_{pin}$, where the pinning scheme is DPP under the rule
``Half $\pm 1$.'' We observe that, in the absence of pinning
control ($\rho_{pin}=0$), herd behavior prevails in the
\emph{free} system, associated with which there are oscillations
with extremely large variances $\sigma^{2}$. Such a fluctuation
state in which $A_t$ oscillates between $0$ and $N$ is in fact an
\emph{absorbing state} of the system, in which the resource
allocation is extremely unreasonable and inefficient. As pinning
control is turned on, even when only a few agents are pinned,
e.g., $\rho_{pin}=0.01$, the fluctuations are weakened
considerably and harmful absorbing state no long exists.
Figure~\ref{fig:At} also shows the corresponding distributions of
$A_t$
for different
cases. The general numerical observation is that pinning control
is highly effective in suppressing or even eliminating herd
behavior.

\section{Mean-field theory of free systems} \label{sec:theory_free_system}

We aim to develop a comprehensive theoretical understanding of the
pinning control method with respect to herd behavior. To gain
insight, we first derive an analytic theory for free systems.

In the mean-field framework, agents in different states are well
mixed. At time step $t$, An individual $i$ of degree $k_i$ has on average
$n^t_{+}=k_i{\rho}^t_{\oplus}$ neighbors that adopt $+1$, where
${\rho}^t_{\oplus}=A_{t}/N$ is the density of $+1$ agents in the
whole system. According to the updating rule Eq.~(\ref{eq:P1}),
$i$ will choose $+1$ at the next time step $t+1$ with the probability
\begin{eqnarray} \label{eqp0}
P_{i\rightarrow\oplus}=k_{i}(1-{\rho}^{t}_{\oplus})/k_{i}=1-{\rho}^{t}_{\oplus}.
\end{eqnarray}
The probability for an agent to choose $-1$ is
$P_{i\rightarrow\ominus}={\rho}^{t}_{\oplus}$. The conditional
transition probability for $A_{t+1}$ agents to select $+1$ at the
next time step $t+1$ obeys the binomial distribution given by
\begin{eqnarray} \label{eq:Pik}
P(A_{t+1}|A_{t}) & = & { N  \choose  A_{t+1}}\cdot
(P_{i\rightarrow\oplus})^{A_{t+1}}\cdot \nonumber\\
&& (1 - P_{i\rightarrow\oplus})^{N-A_{t+1}}.
\end{eqnarray}
The expectation value of $A_{t+1}$ is $E(A_{t+1})=N \cdot
P_{i\rightarrow\oplus}$, and the variance of $A_{t+1}$ about
$E(A_{t+1})$ can be explicitly written as $\delta^2=N \cdot
P_{i\rightarrow\oplus}\cdot(1-P_{i\rightarrow\oplus})$. From
Eq.~(\ref{eqp0}), we have,
\begin{eqnarray}
&&E(A_{t+1})=N \cdot (1-{\rho}^{t}_{\oplus})=N-A_{t}.
\end{eqnarray}
The expected difference of $A_{t+1}$ from the \emph{optimal
solution} $N/2$ is
\begin{eqnarray}
\Delta A_{t+1}&&=  E(A_{t+1})-N/2= N-A_{t}-N/2 \nonumber = -
\Delta A_{t}.
\end{eqnarray}
The relation of the expected departures from the optimal state for
two successive time steps is thus
\begin{eqnarray}
|\Delta A_{t+1}|= | \Delta A_{t}|.\label{eq:deparfree}
\end{eqnarray}
If a large event takes place initially in the system (e.g.,
$A_{t=0}\gg{N/2}$, or $A_{t=0}\ll{N/2}$), the departure from $N/2$
will not decrease, so large oscillations will persist with the
state of the winning side reversing at each time step. In fact,
$A_{t}$ is a Markov-chain process with successive random number
drawn from Eq.~(\ref{eq:Pik}). As soon as $A_{t}$ reaches zero or
$N$ in the stochastic process, $A_{t}$ will oscillate between $0$
and $N$ continuously, landing the free system in an absorbing
state. Herd behavior is thus prevalent in the free system, a
hallmark of which is large oscillations in $A_t$.

A key quantity in the stochastic description of the
resource-allocation process is the distribution $P(A_{t})$, the
probability that $A_{t}$ agents adopt $+1$ at time $t$. Since
$A_{t}$ fluctuates about $N/2$, the choice $+1$ acts as the global
majority and minority choice alternately. The stable distribution
thus obeys $P(A_{t+2l})=P(A_{t})$ and $P(A_{t+2l+1})=P(A_{t+1})$,
for $l\in \mathbf{N}$.

According to Eq.~(\ref{eq:Pik}), the conditional transition
probability for two successive time steps $t$ and $t+1$, we have
the following two-step conditional probability, or the transition
probability:
\begin{eqnarray} \label{eq:T20}
T(A_{t+2},A_{t}) & \equiv & P(A_{t+2}|A_{t}) \\ \nonumber & = &
\sum_{A_{t+1}} P(A_{t+2}|A_{t+1}) \cdot P(A_{t+1}|A_{t}).
\end{eqnarray}
To simplify notation, we set $A_{t}=i$, $A_{t+1}=k$, and
$A_{t+2}=j$, with $i,k,j\in [0,N]$. The conditional transition
probability of the free system is thus given by
\begin{eqnarray}
T(j,i) & \equiv & P(j,t+2|i,t) \nonumber\\
       & =      & \sum_{k} P(j,t+2|k,t+1) \cdot P(k,t+1|i,t) \nonumber\\
       &= & \sum_{k}[ {N  \choose j}\cdot
  (1-\frac{k}{N})^{j}\cdot (\frac{k}{N})^{N-j}] \cdot [{N  \choose k}\cdot \nonumber\\
  & & (1-\frac{i}{N} )^{k}\cdot (\frac{i}{N})^{N-k} ].  \nonumber
\end{eqnarray}
The resulting \emph{balance equation} governing the dynamics of
the Markov chain reads,
\begin{eqnarray} \label{eq:mastereq}
P(j)=\sum_{i}P(j,t+2|i,t)P(i)=\sum_{i}T(j,i)P(i),
\end{eqnarray}
which is in fact a \emph{discrete-time master equation}. For large
$t$, the system evolves into the stable state defined by
$P(i)=P(j)$. Equation~(\ref{eq:mastereq}) can be written in the
matrix form as
\begin{eqnarray} \label{eq:mastereqM}
P(\mathbf{A})=\mathbf{T} P(\mathbf{A}),
\end{eqnarray}
where $\mathbf{T}$ is an $N\times N$ matrix with element
$\mathbf{T}_{ji}=T(j,i)$. The stable distribution of $A_{t}$ is
then $P_1(\mathbf{A})$, the eigenvector of matrix $\mathbf{T}$
associated with eigenvalue $\lambda=1$. For the {\em free} system, we
thus obtain $P(A)=\delta_{A,0}$ or $\delta_{A,N}$ with equal
probability on average, where the exact value of $P(A)$ depends on
the initial condition and the number of time steps (even or odd).
This explains the simulation results in Fig.~\ref{fig:At} for the
case of $\rho_{pin}=0$, where $A_{t}=0,N,0,...$ is an absorbing
state and thus is the stable state of the free system. Once we
obtain the stable distribution $P(\mathbf{A})$ analytically from
Eq.~(\ref{eq:mastereqM}), we can calculate the cumulative variance
[Eq.~(\ref{eq:sigmaM})] of the system as,
\begin{eqnarray} \label{eq:sigmaA}
\sigma^{2}=\frac{\overline{(A_t-N/2)^2}}{1-\rho_{pin}} =
\frac{\sum_{A=0}^{N}P(A)(A-N/2)^2}{1-\rho_{pin}}.
\end{eqnarray}
The fluctuation of the {\em free} system is thus given by
$\sigma^{2}=N^2/4$.

While we have considered the resource-allocation dynamics in
networked systems in which agents interact with each other without
any restriction, the discrete-time master equation
Eq.~(\ref{eq:mastereqM}) can be used to analyze and understand
oscillatory dynamics in general complex adaptive systems.

\section{Mean-field analysis of systems under pinning control}
\label{sec:theory_pinning_control}

We now develop a theory to understand the working of pinning
control in suppressing/eliminating herd behavior. The setting is a
networked system of $N$ agents in which a fraction $\rho_{pin}$ of
the agents are not allowed to choose resources freely. Without
loss of generality, we focus on the ``Half $\pm 1$'' pinning rule.

\subsection{Mean-field analysis for well-mixed free and pinned agents}
\label{subsec:MF_well_mixed}

We first consider the case of random pinning. Under the assumption
that the dynamical properties of pinned and free nodes are
identical, the interactions among them are well-mixed.
Consequently, the probability for the neighbor of one given
\emph{free} agent to be pinned is
\begin{equation} \label{eq:Pfpnode}
P_{p}=N_{p}/N=\rho_{pin},
\end{equation}
where $N_p$ is the number of pinned agents in the system.
For a free agent $i$ with degree $k_i$, the average numbers of
pinned and free neighbors, denoted by $n_{p}$ and $n_{f}$,
respectively, are
\begin{eqnarray} \label{eq:wellmixnfp}
&& n_{f}=(1-P_{p})k_{i}=(1-\rho_{pin})k_{i},\nonumber \\
&& n_{p}=P_{p}k_{i}=\rho_{pin}k_{i},
\end{eqnarray}
where half of a pinned neighbor adopt $+1$, and the other half
adopt $-1$. According to the updating rule Eq.~(\ref{eq:P1}), the
probability for $i$ to choose $+1$ at the next time step $t+1$ is
\begin{equation} \label{eq:Pi+pin}
P_{i\rightarrow\oplus}=\frac{n_{f}(1-{\rho}^{t,f}_{\oplus}) +
n_{p}/2}{k_{i}}=(1-\rho_{pin})(1-{\rho}^{t,f}_{\oplus})+
\frac{\rho_{pin}}{2},
\end{equation}
where ${\rho}_{\oplus}^{t,f}$ stands for the density of free
agents who choose $+1$, and ${\rho}^{t}_{\oplus}$ is the density
of $+1$ agents in the whole system. When $\rho_{pin}=0$, ${\rho}_{\oplus}^{t,f}$ reduced to
${\rho}_{\oplus}^{t}$, we have
$P_{i\rightarrow\oplus}=1-{\rho}^{t}_{\oplus}$, which is reduced
to the result for the free system, that is, Eq.~(\ref{eqp0}).

Using a similar reasoning that leads to the conditional transition
probability of $A_{t+1}$ for the free system as in
Eq.~(\ref{eq:Pik}), we obtain the corresponding result for the
pinning system:
\begin{eqnarray} \label{eq:Pikpin}
&& P(A_{t+1}|A_{t}) = P(A_{t+1}^f|A_{t}^f) \nonumber\\
&& = {N_{f} \choose A_{t+1}^{f}}\cdot
(P_{i\rightarrow\oplus})^{A_{t+1}^{f}}\cdot (1 -
P_{i\rightarrow\oplus})^{N_{f}-A_{t+1}^{f}},
\end{eqnarray}
where $A_{t'}$ and $A_{t'}^{f}$ are related by $A_{t'}=A_{t'}^{f}+N_p/2$, $A_{t'}^{f}$ and $N_p/2$ are the
numbers of free $+1$ agents and pinned $+1$ agents in the system
at time $t'$, respectively. The deviation of $A_{t}$ from the
optimal state $N/2$ is mainly due to the fluctuation of the free agents. From the binomial distribution, we get the
expectation number of the free $+1$ agents at time $t+1$, and the
variance about the expectation number as,
\begin{eqnarray} \label{eq:expectationA}
&&E(A_{t+1}^{f})=N_{f}\cdot P_{i\rightarrow\oplus}, \nonumber\\
&&\delta^2_{f}=N_{f} \cdot
P_{i\rightarrow\oplus}\cdot(1-P_{i\rightarrow\oplus}).
\end{eqnarray}
The expectation number of $+1$ agents (including the pinned $+1$
agents) is
\begin{equation} \label{eq:AandAf}
E(A_{t+1})=E(A_{t+1}^{f})+N_{p}/2,
\end{equation}
which can be written as a function of $\rho_{pin}$ and $A_{t}$:
\begin{eqnarray}
E(A_{t+1}) &&=\underline{(1-\rho_{pin})}\cdot (N-A_{t}) +
\underline{\rho_{pin}}\cdot \frac{N}{2}, \label{eq:meanA1} \\
&&=(N-A_{t})+\rho_{pin}\cdot(A_{t}-\frac{N}{2}). \label{eq:meanA3}
\end{eqnarray}
From Eq.~(\ref{eq:meanA3}), we obtain the following expected
difference from $N/2$:
\begin{eqnarray}
\Delta A_{t+1}&&=  E(A_{t+1})-N/2=
-(1-\rho_{pin})\cdot(A_{t}-N/2) \nonumber \\
&&= -(1-\rho_{pin})\cdot \Delta A_{t}
\end{eqnarray}
The relation between the expected deviations from the optimal
state $N/2$ for two successive time steps is then given by
\begin{eqnarray} \label{eq:deparpin}
|\Delta A_{t+1}|= (1-\rho_{pin})\cdot | \Delta A_{t}|.
\end{eqnarray}
Compared with the expected departure obtained in the free system,
as given by Eq.~(\ref{eq:deparfree}), we see that, once pinning is
implemented, the deviation from the optimal state decays by the
factor $(1-\rho_{pin})$ at each time step and, consequently, the
oscillation of the system is suppressed. In case of large events,
pinning will make $A_{t}$ to approach the equilibrium value $N/2$.

\begin{figure}
\begin{center}
\epsfig{figure=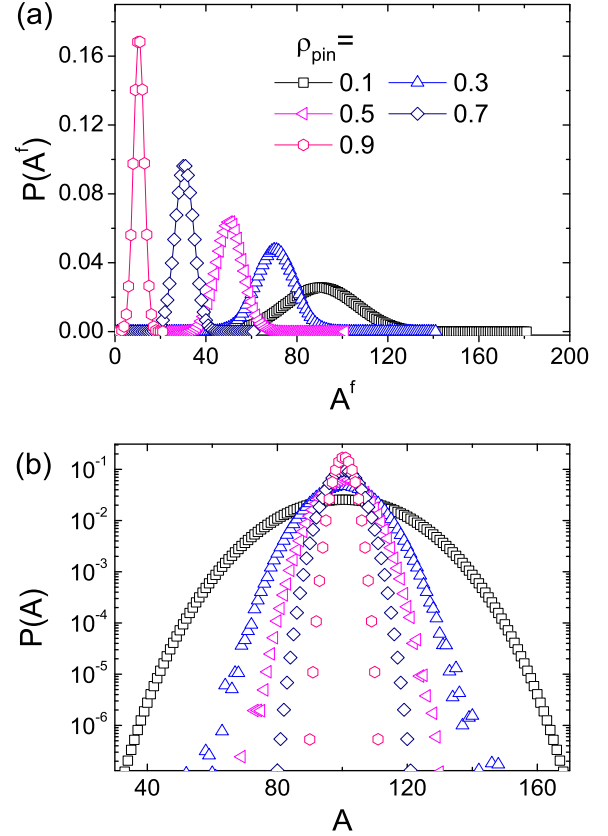,width=\linewidth} \caption{(Color
online). Analytical results of the stable distribution of $A^f$
(a) and $A$ (b) for a system of $N=201$ agents, from the
mean-field analysis under the assumption of well-mixed
interactions between the pinned and free agents, as given by
Eq.~(\ref{eq:T20pin1}). The graph $P(A)$ in (b) is on a
logarithmic-normal plot.} \label{fig:P(A)}
\end{center}
\end{figure}

A stochastic analysis similar to that for the free system can then
be carried out for a pinning system. From Eqs.~(\ref{eq:T20}) and
(\ref{eq:Pikpin}), we can get the conditional transition
probability from time step $t$ to $t+2$ as,
\begin{eqnarray} \label{eq:T20pin1}
&& T(j,i)\equiv P(j,t+2|i,t) \\
&&=\sum_{k} P(j,t+2|k,t+1) \cdot P(k,t+1|i,t) \nonumber\\ =
&&\sum_{k}\{ [{N_f\choose j-\frac{1}{2}N_{p}}\cdot (1-\frac{k}{N}
)^{j-\frac{1}{2}N_{p}}\cdot
(\frac{k}{N})^{N_{f}-(j-\frac{1}{2}N_{p})} ] \nonumber\\
&&\cdot [ {N_f\choose k-\frac{1}{2}N_{p}} \cdot (1-\frac{i}{N}
)^{k-\frac{1}{2}N_{p}}\cdot
(\frac{i}{N})^{N_{f}-(k-\frac{1}{2}N_{p})} ] \}, \nonumber
\end{eqnarray}
where $i,k,j\in [N_{p}/2,N-N_{p}/2]$, which denote $A_{t}$,
$A_{t+1}$, and $A_{t+2}$, respectively. Following the steps from
Eq.~(\ref{eq:mastereq}) to Eq.~(\ref{eq:sigmaA}) for a free
system, we can derive formulas of $P(A^f)$ and $P(A)$ for the
pinning system, based on the assumption Eq.~(\ref{eq:Pfpnode})
that the pinned and free nodes are identical with well-mixed
interactions. Figure~\ref{fig:P(A)} shows the corresponding
results for different values of $\rho_{pin}$. We see that the
stable distribution $P(A)$ has a Gaussian profile, with the
expectation value of $E(A)=N/2$, which should be compared with the
value $P(A)=\delta_{A,0}$ or $\delta_{A,N}$ for the free system.
This result indicates that the harmful absorbing state associated
with a free system has essentially been eliminated even when only
a few agents are pinned. Representative numerical evidence
supporting this result is shown in Fig.~\ref{fig:At} for FCN and
SFN with only $1\%$ of the agents pinned.

\begin{figure}
\begin{center}
\epsfig{figure=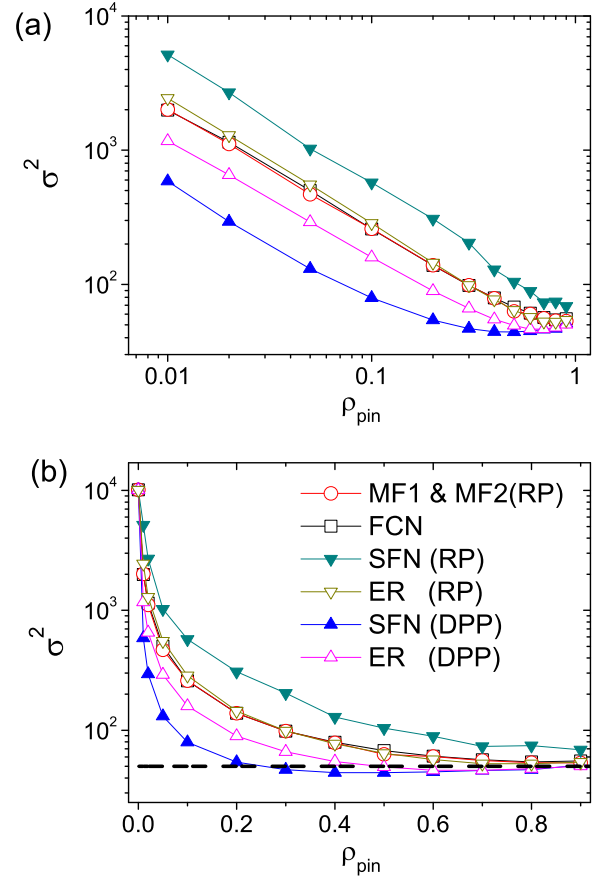,width=\linewidth} \caption{(Color
online). Modified cumulative variance $\sigma^2$
[Eq.~(\ref{eq:sigmaM})] as a function of $\rho_{pin}$ from
mean-field analysis (red circle marked by ``MF1''), where pinned
and free nodes are well mixed as in Eq.~(\ref{eq:T20pin1}).
Simulation results for FCN, SFN, and ER random network are also
shown, where all network size is $N=201$, the pinning schemes are
RP and DPP under the ``Half $\pm 1$'' rule, the average degrees of
SFN and ER random networks is $\langle k\rangle=6$, and the number
of network realizations is $200$. The graphs in (a) and (b) are on
a logarithmic and a logarithmic-linear scale, respectively.}
\label{fig:MF1Simu}
\end{center}
\end{figure}

From Eq.~(\ref{eq:sigmaA}), we can calculate the variance
$\sigma^2$ of the system for different values of $\rho_{pin}$
analytically, as shown by the open circle marked ``MF1'' in
Fig.~\ref{fig:MF1Simu}. Simulation results for FCN (open black
square), SFN (solid triangle) and ER random network (open
triangle) under DPP (up triangle) or RP (down triangle) and the
``Half $\pm 1$'' rule are also shown. We observe that the variance
$\sigma^2$ of the system decreases dramatically in a power-law
manner as pinning control is turned on, and the agreement between
theoretical prediction and numerical simulations for FCN is good.
However, for ER random network and SFN, there is marked difference
between the theoretical and numerical results, especially for the
DPP scheme, indicating that the approach of mean-field, stochastic
type of analysis may not be adequate to account for the behavior
of the system under pinning control. In the following, we shall
develop a modified mean-field analysis to overcome this
difficulty.

\subsection{Modified mean-field analysis}
\label{subsec:MF_modified}

The assumption Eq.~(\ref{eq:Pfpnode}) in which free and pinned
agents are identical and well mixed may not be valid in general,
especially when the underlying network is heterogeneous, such as
SFNs. In such a case, the probability for a free node to contact
with a pinned node will deviate from $\rho_{pin}$, requiring
modifications of the conventional mean-field analysis.

\subsubsection{Analysis of degree-preferential pinning
on scale-free networks}

We first discuss the DPP scheme on SFNs generated by the classic
preferential-attachment rule \cite{BA:1999}, with the degree
distribution given by $P(k)={2m^2}/{k^3}$, where $m$ is the number
of edges each new node brings in as the system grows. The average
degree of the network is $\langle k\rangle=2m$, and the minimum
degree is $k_{min}=m$. For DPP scheme from large to small degree,
the density of pinned agents $\rho_{pin}$ and the minimum degree
of pinned agents (denoted by $k'$) are related to each other as
\begin{eqnarray}
\rho_{pin}=\int_{k'}^\infty P(k)dk,
\end{eqnarray}
giving
\begin{equation} \label{eq:rho-k'}
k'=\sqrt{\frac{m^2}{\rho_{pin}}},
\end{equation}
which can be used to distinguish pinned and free agents in terms
of their degrees, i.e., an agent with $k\geq k'$ (or  $k<k'$) is
pinned (or free). The total number of links in the whole network,
denoted by $L$, is
\begin{eqnarray} \label{eq:L}
L=\frac{1}{2}\int_{k_{min}}^\infty kNP(k)dk.
\end{eqnarray}
The number of the so-called pinning-affected links $L_{pin}$ and
that of free-related links $L_{free}$ can be defined,
respectively, as
\begin{eqnarray}
&&L_{pin}=\frac{1}{2}\int_{k'}^\infty kNP(k)dk,\\
&&L_{free}=\frac{1}{2}\int_{k_{min}}^{k'}kNP(k)dk,
\end{eqnarray}
where $L_{pin}+L_{free}=L$. From the fraction of pinning-affected
links, we have the following probability for one neighbor of a
given free agent to be a pinned agent:
\begin{equation} \label{eq:Pfplink}
P_{p}= L_{pin}/L.
\end{equation}
For a SFN under DPP, we have $L=mN$,
$L_{pin}=mN\sqrt{\rho_{pin}}$, and
$L_{free}=mN(1-\sqrt{\rho_{pin}})$ and, consequently,
$P_{p}=\sqrt{\rho_{pin}}$. It should be noted that
Eq.~(\ref{eq:Pfplink}) differs from Eq.~(\ref{eq:Pfpnode}) in that
the former is expressed in terms of the pinning-affected links and
the latter is with respect to the fraction of pinned agents. This
difference underlies our modified mean-field analysis.

\begin{figure}
\begin{center}
\epsfig{figure=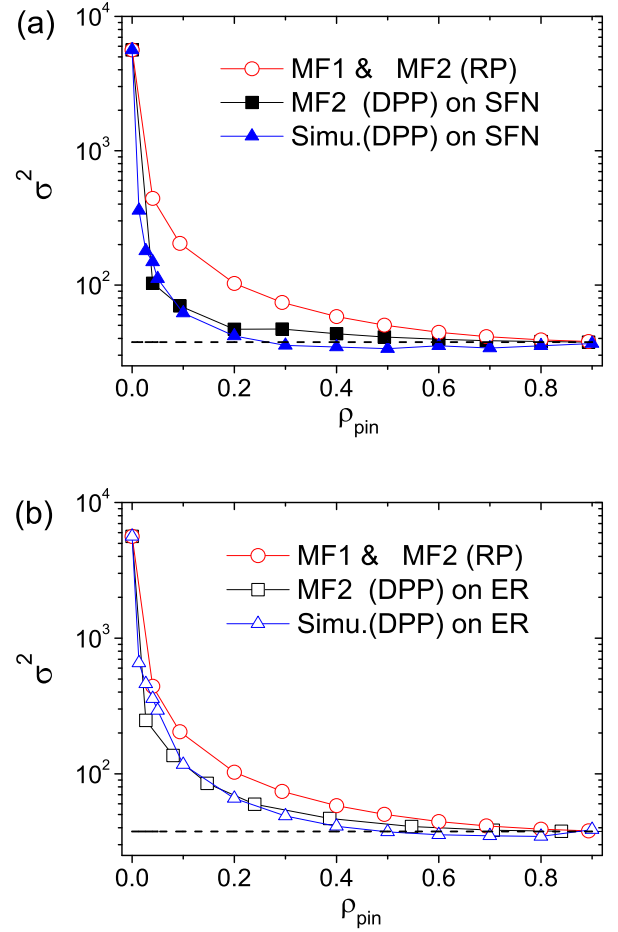,width=\linewidth} \caption{(Color
online). Modified cumulative variance $\sigma^2$ as a function of
$\rho_{pin}$ as predicted by the mean-field analysis (red circle
marked by ``MF1''), by our modified mean-field analysis (black
square marked by ``MF2''), in comparison with the simulation
results (blue triangle). Note that ``MF2'' under RP scheme is the same as ``MF1'' (see Sec. \ref{sec:MF2RP}). Simulations are for SFNs (a) and ER
random networks (b), all of average degree 6.  Network size is
$N=150$ and the pinning scheme is DPP under the ``Half $\pm 1$''
rule. The simulation results are averaged over $1000$
realizations.} \label{fig:MF2Simu}
\end{center}
\end{figure}

Utilizing Eq.~(\ref{eq:Pfplink}), we can write the average numbers
of the free and pinned neighbors for a free agent $i$ of degree
$k_i$ as
\begin{eqnarray}
&& n_{f}=(1-P_{p})k_{i}=(1-\sqrt{\rho_{pin}})k_{i},\\
&& n_{p}=P_{p}k_{i}=\sqrt{\rho_{pin}}
\end{eqnarray}
Similar to the analysis procedure from Eq.~(\ref{eq:Pi+pin}) to
Eq.~(\ref{eq:T20pin1}), we can obtain the corresponding results
for SFNs with DPP under the ``Half $\pm 1$'' rule. The probability
for agent $i$ to choose $+1$ is
\begin{eqnarray} \label{eq:Pi+pin2}
P_{i\rightarrow\oplus}&&=\frac{n_{f}(1-{\rho}^{t,f}_{\oplus})+n_{p}/2}{k_{i}} \nonumber\\
&&=(1-P_{p})(1-{\rho}^{t,f}_{\oplus})+P_{p}/2 \nonumber\\
&&=\frac{1-P_{p}}{1-\rho_{pin}} (1-{\rho}^{t}_{\oplus}) +
\frac{P_{p}-\rho_{pin}}{2(1-\rho_{pin})}\nonumber \\
&&\equiv a(1-{\rho}^{t}_{\oplus})+b.
\end{eqnarray}
The expectation numbers of free $+1$ agents and all $+1$ agents
[$E(A_{t+1}^{f})$ and $E(A_{t+1})$, respectively] can then be
obtained from Eqs.~(\ref{eq:expectationA}) and (\ref{eq:AandAf}).
The expected deviation of $A_{t}$ from the optimal state for two
successive time steps is given by
\begin{eqnarray} \label{eq:deparpin2}
|\Delta A_{t+1}|= (1- \sqrt{\rho_{pin}})\cdot | \Delta A_{t}|.
\end{eqnarray}
Furthermore, from Eqs.~(\ref{eq:Pikpin}) and (\ref{eq:Pi+pin2}),
we can get the conditional transition probability from time step
$t$ to $t+2$ as
\begin{eqnarray} \label{eq:T20pin2}
&& T(j,i)\equiv P(j,t+2|i,t) \nonumber\\
&& =\sum_{k} P(j,t+2|k,t+1) \cdot P(k,t+1|i,t) \nonumber\\
&& = \sum_{k}\{ [{N_f\choose j-\frac{1}{2}N_{p}}\cdot
(\frac{P_{p}}{2}+P_{f}\cdot\frac{N_{f}-k+\frac{N_{p}}{2}}{N_{f}})^{j-\frac{1}{2}N_{p}}
\cdot \nonumber\\
&&
(\frac{P_{p}}{2}+P_{f}\frac{k-\frac{N_{p}}{2}}{N_f})^{N_{f}-(j-\frac{1}{2}N_{p})}
]
\nonumber\\
&&\cdot [ {N_f\choose k-\frac{1}{2}N_{p}} \cdot
(\frac{P_{p}}{2}+P_{f}\cdot\frac{N_{f}-i+\frac{N_{p}}{2}}{N_{f}}
)^{k-\frac{1}{2}N_{p}}
\cdot \nonumber\\
&&
(\frac{P_{p}}{2}+P_{f}\frac{k-\frac{N_{p}}{2}}{N_f})^{N_{f}-(i-\frac{1}{2}N_{p})}
]\},
\end{eqnarray}
where $i,k,j\in [N_{p}/2,N-N_{p}/2]$ are associated with $A_{t}$,
$A_{t+1}$, and $A_{t+2}$, respectively, and $P_{f}=1-P_{p}$.
Similar to analysis of free systems [Eqs.~(\ref{eq:mastereqM}) and
(\ref{eq:sigmaA})], we obtain the analytic results of
$P(\mathbf{A})$ and $\sigma^2$ of the pinning system in terms of
the fraction of pinning-affected links, as shown in
Fig.~\ref{fig:MF2Simu} [marked by ``MF2 (DPP) on SFN''], together
with the corresponding simulation results [marked by ``Simu.(DPP)
on SFN'']. For comparison, result from the original mean-field
analysis (MF1) is also included. We see that our modified
mean-field analysis yields results that match more closely those
from simulations.

\subsubsection{Degree-preferential pinning on random networks}

The degree of ER random network \cite{ER:1960} obeys Poisson
distribution:
\begin{equation}
P(k)=\frac{e^{-\langle k\rangle}\langle k\rangle^{k}}{k!}.
\end{equation}
The relation between $k'$ and $\rho_{pin}$ can then be written as
\begin{equation}
\rho_{pin}=\sum_{k=k'+1}^{k_{max}}P(k),
\end{equation}
where the maximum degree $k_{max}$ for a network of size $N$ can
be calculated by $P(k_{max})\approx 1/N$. The degree $k'$ for a
given $\rho_{pin}$ can be calculated numerically. The quantities
$L$, $L_{pin}$, and $L_{free}$ are, respectively, given by
\begin{eqnarray}
&&L=\frac{1}{2}\sum_{k=1}^{\infty}kNP(k) =
\frac{1}{2}\sum_{k=1}^{k_{max}}
\frac{kNe^{-\langle k\rangle}\langle k\rangle^k}{k!} \\
&&L_{pin}=\frac{1}{2}\sum_{k=k'+1}^{\infty}kNP(k) \nonumber\\
&& = \frac{1}{2}\sum_{k=k'+1}^{k_{max}}
\frac{kNe^{-\langle k\rangle}\langle k\rangle^k}{k!} \\
&&L_{free}=\frac{1}{2}\sum_{k=1}^{k'}kNP(k)=\frac{1}{2}
\sum_{k=1}^{k'} \frac{kNe^{-\langle k\rangle}\langle
k\rangle^k}{k!}
\end{eqnarray}
Following a similar modified mean-field analysis for SFNs, we can
calculate $k'$ for a given value of $\rho_{pin}$. The quantities
$P_{p}$, $P_{i\rightarrow\oplus}$, $T(j,i)$, the stable
distribution $P(\mathbf{A})$, and finally $\sigma^2$ can then be
obtained as a function of $\rho_{pin}$, as shown in
Fig.~\ref{fig:MF2Simu}(b). We see that the modified mean-field
analysis [marked by ``MF2 (DPP) on ER''] gives more accurate
prediction about the system behaviors.

\subsubsection{Random pinning}\label{sec:MF2RP}

For random pinning on a network of a given degree distribution
$P(k)$, the number of pinning-affected links and free links are
\begin{eqnarray}
&& L_{pin} = \frac{1}{2}\int_{k_{min}}^\infty kN\rho_{pin}P(k)dk = \rho_{pin}L,\\
&& L_{free} = \frac{1}{2}\int_{k_{min}}^{\infty}kN(1-\rho_{pin})P(k)dk \nonumber\\
&& = (1-\rho_{pin})L,
\end{eqnarray}
where the number $L$ of total links is given by Eq.~(\ref{eq:L}).
In the RP process, the value of $L_{pin}$ and $L_{free}$ are
independent of the degree distribution $P(k)$. We thus have
\begin{equation}
P_{p}= L_{pin}/L= \rho_{pin}.
\end{equation}
Similar to the analysis of DPP on heterogeneous networks, we can
obtain $n_f$ and $n_p$ and substitute them into
Eq.~(\ref{eq:Pi+pin2}) to get
$P_{i\rightarrow\oplus}=1-{\rho}^{t}_{\oplus}$. We see that the
quantities $P_{p}$ and $P_{i\rightarrow\oplus}$ for RP are the
same as those given by Eqs.~(\ref{eq:Pi+pin}) and
(\ref{eq:Pi+pin2}) from the idealized mean-field analysis,
regardless of the network structure. The reason is that for RP,
the pinned and free nodes tend to mix well on the network,
satisfying the basic mean-field assumption. A consequence is then
that the relation between $\Delta A_{t+1}$ and $\Delta A_{t}$, the
conditional transition probability $T(j,i)$, and the stable
distribution $P(\mathbf{A})$ are identical to those given by the
idealized mean-field analysis [Eq.~(\ref{eq:Pfpnode}) to
Eq.~(\ref{eq:T20pin1})]. As a consequence, the analytical results
for random pinning from the modified mean-field analysis [marked
by ``MF2 (RP)''] are the same as those from the original
mean-field analysis [marked by ``MF1''], as shown in
Figs.~\ref{fig:MF1Simu} and \ref{fig:MF2Simu}.

\section{Conclusion and Discussion} \label{sec:conclusion}

The collective behavior of herding can occur commonly in complex
resource-distribution systems, the hallmark of which is strong and
even extreme fluctuations in the usage of available resources. In
particular, for a free system without any external intervention,
typically the resources are accessed and utilized in a highly
non-uniform manner: there are time intervals in which almost all
resources are used, followed by those in which most agents in the
system focus on only a few resources. Such an uneven utilization
of resources makes the system inefficient and is generally
harmful. What we have shown in this paper is that, implementing a
simple pinning control scheme can effectively eliminate herding.
While the idea of pinning control has been used widely to control
complex networked systems
\cite{WC:2002,LWC:2004,CLL:2007,XLCCY:2007,TWF:2009,PF:2009,YCL:2009},
our contribution is to introduce it to complex resource-allocation
systems. More importantly, we have developed a solid physical
theory based on the mean-field approach and its variant to
establish the theoretical foundation of the pinning control in
such systems. Specifically, we have analyzed the approaches of
random and degree-preferential pinning on networks of distinct
topologies, and demonstrated that a non-random type of control
strategy can be more effective than a random one [cf.
Figs.~\ref{fig:MF1Simu} and \ref{fig:MF2Simu}]. The basic philosophy underlying our control scheme is ``to pin a few to benefit the majority.'' That is, fixing a few agents' choice of resource utilization can reduce significantly the fluctuations in the whole system, resulting in remarkable improvement in its efficiency.

Our theory suggests that the best strategy to reduce fluctuations
through pinning is to choose the agents of high degrees. However,
one difficulty associated with the degree preferential pinning
scheme is that it requires fairly complete knowledge about the
degree of each agent in the network. This is especially
challenging for real-world networks, where global information
about the network may not be available to every agent. In
addition, the interactions among the agents when competing for
resources may not be readily quantified. An important issue
concerns thus how herd behavior can be controlled when information
about the network structure and interactions among the agents is
lacking. Immunization method developed in controlling virus
spreading on complex networks \cite{CHbA:2003}, which requires no
detailed knowledge about the network and its interacting dynamics,
may provide a viable approach. For example, one can consider the
scheme of acquaintance pinning, in which random acquaintances of
random nodes are pinned in their selection of resources.

Real-world systems for which Boolean game model and pinning scheme
may be applicable include the financial market systems, urban
traffic systems, computer network systems, and so on. In these
systems, individuals' choice can be ``pinned'' by means of certain
incentive policies with compensations or rewards. The incentive
policy for pinning can be modelled as random fields in the
dynamics and may introduce a cost linear to the number of pinned
agents. We see that the system welfare, i.e., the performance of
the resource allocation system measured by the variance of the
number of agents choosing a resource, improves rapidly as soon as
very few pinnings take place. Take the financial market system as
an example, where the policies of the Market Makers are the
strategy to intervene the game dynamics in the market by certain
regulations or incentives so as to make the capital allocation
more efficient, i.e., to realize the goal of achieving efficient
markets. Our study of pinning control is directly relevant to
these real-world examples. In addition to its real significance,
our work represents a basic and general mathematical framework to
address the role of pinning in complex resource-allocation
dynamics in social, economical and political systems.

\section*{Acknowledgement}

This work was partially supported by the NSF of China under Grants
No. 11275003, 11135001, and 10905026. Z.G.H. was supported by the
Lanzhou DSTP No. 2010-1-129. Y.C.L. was supported by AFOSR under
Grant No. FA9550-10-1-0083.

\end{document}